\documentstyle[preprint,aps]{revtex}

\tightenlines
\input amssym.def
\input amssym.tex
\newcommand{\beq}{\begin{equation}}
\newcommand{\eeq}{\end{equation}}
\newcommand{\eps}{\epsilon}
\newcommand{\z}{\zeta}
\newcommand{\lam}{\lambda}
\newcommand{\bml}{\begin{eqnarray}}
\newcommand{\eml}{\end{eqnarray}}

\begin{document}

\title{The Specific Heat of a Trapped Fermi Gas: \\ an Analytical
Approach}

\author{Jos\'{e} M. B.
Noronha\footnote{Corresponding author. E-mail:J.M.B.D.Noronha@ncl.ac.uk}
\and David J.
Toms\footnote{E-mail:D.J.Toms@ncl.ac.uk}}

\address{Department of Physics, University of Newcastle upon Tyne, \\
Newcastle upon Tyne, NE1 7RU, U. K.}

\date{\today}

\maketitle

\begin{abstract}
We find an analytical expression for the specific heat of
a Fermi gas in a harmonic trap using a semi-classical
approximation. Our approximation is valid for $k_{B}T
\gtrsim \hbar w_{x,y,z}$ and in this range it is
shown to be highly
accurate. We comment on the
semi-classical approximation, presenting an explanation for this high
accuracy.  

\end{abstract}


\section*{}

Trapped ultra-cold Fermi gases have received a lot of attention in the
last few years. This was partly triggered by the first observations of
Bose-Einstein condensation in 1995 \cite{1,2,3}. Since then, the field has
vastly expanded. The trapping and cooling of fermions is in a much less
developed state than that of bosons but it will probably not be long
before quantum degeneracy is achieved \cite{4}. (For recent experiments
see \cite{5,6}.)

On the theoretical side, Butts and
Rokhsar \cite{7} have calculated the specific heat numerically and
provided analytical results
for the spatial and momentum distribution of a non-self-interacting
spin polarised Fermi gas at $T=0$
in a harmonic trap using the semi-classical (Thomas-Fermi) approximation.
Their results are
valid for large particle numbers. In addition, Schneider and Wallis
\cite{8} made a
numerical study of a similar gas and focused on the effects of small
particle numbers. The present paper complements that of Butts and Rokhsar
in that we provide an analytical expression for
the specific heat of a non-self-interacting spin polarised Fermi gas in
the context of a
semi-classical approximation. We then compare it to the exact result given
by a numerical
calculation. As we shall see, the semi-classical approximation produces
extremely good
results for $kT\gtrsim \hbar w_{x,y,z}$, where $w_{x,y,z}$ are the
frequencies of the trap in the three spatial directions, for particle
numbers as low as $N=1000$. This high level of accuracy is perhaps
unexpected. We will discuss this and an explanation will be given.
The characteristic temperature for quantum
degeneracy, the Fermi temperature, is given by $kT_{F}=\hbar
w(6N)^{\frac{1}{3}}$ for an isotropic trap. The condition we impose is 
therefore valid for a quantum
degenerate regime. It is also worth noting that this condition holds in
current experiments with trapped fermions \cite{6}.

The effects of interactions has been dealt with in several works
\cite{9,10,11,12,13,14,15,16} and in particular the possibility of a BCS
transition has been
studied in the case of $^{6}$Li \cite{9,10,11}.
Considering the gas to be non-interacting is a very good approximation for
a dilute neutral atomic gas. The only case where interactions could be
important is trapped $^{6}$Li with two hyper-fine spin states \cite{12}.
$^{6}$Li
has an anomalously large s-wave scattering length. At least two different
states are needed for s-wave scattering as it is forbidden for particles
in the same state. However, the results of ref.\ \cite{11} are for
$kT<<\hbar w$,
lying outside the range of temperatures concerned in the present work.

Consider a gas of fermions in a harmonic potential. Neglecting
interactions between the particles, the energy levels of each particle are
\bml
E_{n_{1}n_{2}n_{3}}&=&(n_{1}+\frac{1}{2})\hbar 
w_{1}+(n_{2}+\frac{1}{2})\hbar w_{2}+(n_{3}+\frac{1}{2})\hbar w_{3} \; ,
\\ n_{i}&=&0,1,2,\ldots \; , \nonumber
\eml
where $w_{1,2,3}$ are the frequencies of the trap.

We will use grand--canonical statistics throughout. This makes the
calculations much easier and is justified as the difference between
canonical and grand-canonical results is minute for the particle
numbers we consider \cite{8}.

The number of particles in the system is 
\beq
N=\sum_{n} [e^{\beta (E_{n}-\mu)}+1]^{-1} \; .
\eeq
The internal energy of the system is given by
\beq
U=\sum_{n} E_{n}[e^{\beta (E_{n}-\mu)}+1]^{-1} \; .
\eeq
$\mu $ is the chemical potential and the sum in (2) and (3) is over all 
particle states.
Here it is convenient for simplicity of the
expression, to make the substitutions
\bml
w_{1}&=&w \; , \nonumber \\
w_{2}&=&\lam w \; , \nonumber \\
w_{3}&=&\lam ' w \; , \nonumber \\
x&=&\beta \hbar w \; , \nonumber \\
\mu &=&\hbar w\left( \frac{1+\lam +\lam '}{2}-\epsilon \right) \; ,
\eml
where $x$, $\epsilon $, $\lam $ and $\lam '$ are newly defined dimensionless
variables.
The internal energy becomes
\beq
U=\frac{1}{2}\hbar w(1+\lam +\lam ')N+\frac{1}{2}\hbar w u \; ,
\eeq
where
\bml
u&=&2\sum_{n_{1}=0}^{\infty }\sum_{n_{2}=0}^{\infty }\sum_{n_{3}=0}^{\infty }
k[e^{x(k+\epsilon )}+1]^{-1} \; ,  \\
k&=&n_{1}+\lam n_{2}+\lam 'n_{3} \; . \nonumber
\eml
Only the second term
contributes to the specific heat since $N$ is held fixed. We have
\beq
C
=\left( \frac{\partial U}{\partial T}\right) _{N,w}=\frac{1}{2}\hbar
w\left( \frac{\partial u}{\partial T}\right) _{N,w} \; .
\eeq
Note that the partial derivative at constant particle number and trap
frequency is the natural way of defining the specific heat for a gas in
the situation concerned. After a change of variables from $T$ to $x$
this becomes
\beq
\frac{C}{k_{B}}=-\frac{1}{2}x^{2}\left( \frac{\partial u}{\partial
x}\right)
_{N,w} \; ,
\eeq
where $k_{B}$ is Boltzmann's constant.

The simplest approximation consists in replacing
the triple sums in (6) by a triple integral. Thus,
\beq
u\simeq v=2\int_{0}^{+\infty }\int_{0}^{+\infty }\int_{0}^{+\infty }
k[e^{x(k+\eps )}+1]^{-1}dy_{1}dy_{2}dy_{3}
\eeq
and $k$ is now the real function $k=y_{1}+\lam y_{2}+\lam 'y_{3}$. Changing
the integrating variable to $k$ we have
\beq
v=\frac{1}{\lam \lam '}\int_{0}^{+\infty }k^{3}[e^{x(k+\eps )}+1]^{-1}dk
\; .
\eeq

Note that for the isotropic case we  could have transformed the triple sum
in (6) into a single sum and only then convert it to an integral, yielding
\beq
v=\frac{1}{\lam \lam '}\int_{0}^{+\infty }(k^{3}+3k^{2}+2k)
[e^{x(k+\eps )}+1]^{-1}dk\; ,
\eeq
which, at first sight, might be thought to be a more accurate expression.
For the
anisotropic case, though not trivial, we could likewise have a density of
states expanded to three terms \cite{17,18}. We will coment on this
shortly.

To solve the integral in (10) we Taylor expand the integrand in powers of
$e^{x(k+\eps )}$ and integrate each resulting term. Here we have to
consider two different cases: $\eps >0$ and $\eps <0$. For $\eps >0$ this
yields
\beq
v= \frac{6}{\lam \lam 'x^{4}}\sum_{n=1}^{\infty
}(-1)^{n+1}\frac{e^{-nx\eps }}{n^{4}}
\; .
\eeq
From (8) and (12) and retaining the fact that $v$ is an approximated
version of $u$, we have
\beq
\frac{C}{k_{B}}\simeq \frac{12}{\lam \lam 'x^{3}}A_{4}+\frac{3}{\lam 
\lam 'x^{2}}
\left( \frac{\partial (\eps x)}{\partial x}\right) _{N,w}A_{3} \; ,
\eeq
where $A_{i}$ denotes the sum
\beq
A_{i}=\sum_{n=1}^{\infty }(-1)^{n+1}\frac{e^{-nx|\eps |}}{n^{i}} \; .
\eeq
The use of the absolute value of $\eps $ is redundant in the case we are
considering ($\eps >0$) but it is useful when we consider the $\eps <0$
case.

To obtain an expression for the partial derivative in (13) we use the
equality
\beq
\left( \frac{\partial N}{\partial x}\right) _{N,w}=0 \; ,
\eeq
with $N$, the number of particles, being
\beq
N\simeq \frac{A_{3}}{\lam \lam 'x^{3}} \; .
\eeq
Expression (16) is obtained using the same process as was used to obtain
(13). We then have
\beq
\left( \frac{\partial (\eps x)}{\partial x}\right)
_{N,w}\simeq -\frac{3}{x}\frac{A_{3}}{A_{2}}
\eeq
and
\beq
\frac{C}{k_{B}}\simeq \frac{3}{\lam \lam 'x^{3}}\left(
4A_{4}-3\frac{A_{3}^{2}}{A_{2}}\right) \; .
\eeq
Finally we note that the sums $A_{i}$ can be put in terms of
polylogarithms \cite{19}, yielding
\beq
\lam \lam '\frac{C}{k_{B}}\simeq 3x^{-3}\left[ 4Li_{4}(e^{-x\eps
})-\frac{1}{2}Li_{4}(e^{-2x\eps })-3\frac{[Li_{3}(e^{-x\eps
})-\frac{1}{4}Li_{3}(e^{-2x\eps })]^{2}}{Li_{2}(e^{-x\eps
})-\frac{1}{2}Li_{2}(e^{-2x\eps })}\right] \; .
\eeq

For $\eps <0$, the procedure is similar but slightly more complicated. The
integral of expression (10) must be divided in two parts, according to
whether the exponential is greater or smaller than $1$, as the Taylor
expansion of the integrand is different for the two cases. The final
result for $\eps <0$ is
\beq
\lam \lam '\frac{C}{k_{B}}\simeq -12A_{4}x^{-3}+21\z (4)x^{-3}+6\z (2)\eps
^{2}x^{-1}+\frac{1}{2}\eps ^{4}x+\frac{(A_{3}-\z (2)\eps x-\frac{1}{6}\eps
^{3}x^{3})^{2}}{A_{2}-\z (2) -\frac{1}{2}\eps ^{2}x^{2}} \; ,
\eeq
where $\z $ is the Riemann zeta function. Putting the sums $A_{i}$ in
terms of polylogarithms we have
\bml
\lam \lam '\frac{C}{k_{B}}&\simeq &[-12Li_{4}(e^{x\eps
})+\frac{3}{2}Li_{4}(e^{2x\eps })+21\z (4)]x^{-3}+6\z (2)\eps ^{2}x^{-1}
+\frac{1}{2}\eps ^{4}x+ \nonumber \\ & &
\frac{[Li_{3}(e^{x\eps
})-\frac{1}{4}Li_{3}(e^{2x\eps })-\z (2)\eps x-\frac{1}{6}\eps ^{3}x^{3}]^{2}}
{Li_{2}(e^{x\eps})-\frac{1}{2}Li_{2}(e^{2x\eps })-\z (2)-\frac{1}{2}\eps
^{2}x^{2}} \; .
\eml

Note that these analytical expressions are not an approximation over the
Thomas-Fermi result. They are the exact Thomas-Fermi result, which of
course, is in itself an approximation. We computed the specific heat
numerically using an exact expression
composed of several sums, which is easily obtained inserting (6) in (8)
and
using (15) to get an expression for $(\partial (\eps x)/\partial
x)_{N,w}$.
This is then compared to the values for the specific heat given by (18)
and
(20). The accuracy of the Thomas-Fermi approximation is extremely good for
$x$ of
order 1 or less, as can be seen in figures 1 and 2. As an example,
for $N=1000$ and $x_{1}=x_{2}=x_{3}<2.5$ the error in $C$ is always less
than $1\% $ and if we take $x_{i}<1$, the error is always less than 
$0.1\% $. For larger values of $x$ the approximation rapidly
deteriorates. The results for a higher number of particles are even more
accurate, as expected from this kind of approximation \cite{7}.

Given that for this approximation we have replaced triple sums by triple
integrals, such a level of accuracy is quite surprising. Note that the
expression inside the sums in (6) falls off exponentially.
Therefore the
first few terms are very important and expression (9) should differ
significantly from (6). And indeed it does. The reason our results are so
accurate is that the value of $\eps $ used in (18) and (20) is not the
same as the
one used in (6). For our Thomas-Fermi approximation we took $\eps $ from
expression (16), ie, an approximate $\eps $. So, the insertion of an
approximate $\eps $ in an approximate expression leads to a fortunate
cancelation of errors. We have tested this by inserting the exact value of
$\eps $ into (18) and (20). This leads to much worse results.
Also, we have tried to further improve our approximation by considering the
first two terms of the cubic polynomial in (11) instead of only the first
term and doing the same to the number of particles, obtaining an additional
term in (16) as well as additional terms in (18) and (20). To our
surprise, this only
increases the error. We can only assume that the very high level of
accuracy of the Thomas-Fermi approximation is a happy coincidence. In fact,
if we take the exact value of $\eps $ and input it in expressions (18) and
(20) and in the equivalent expressions obtained using $k^{3}+3k^{2}$ in
(11) instead of only $k^{3}$, then the latter yield the most accurate
result, though still not a very good one.

\acknowledgements

J.N. acknowledges financial support by the Portuguese Foundation for
Science and Technology under grant 
PraxisXXI/ BD/5660/95.

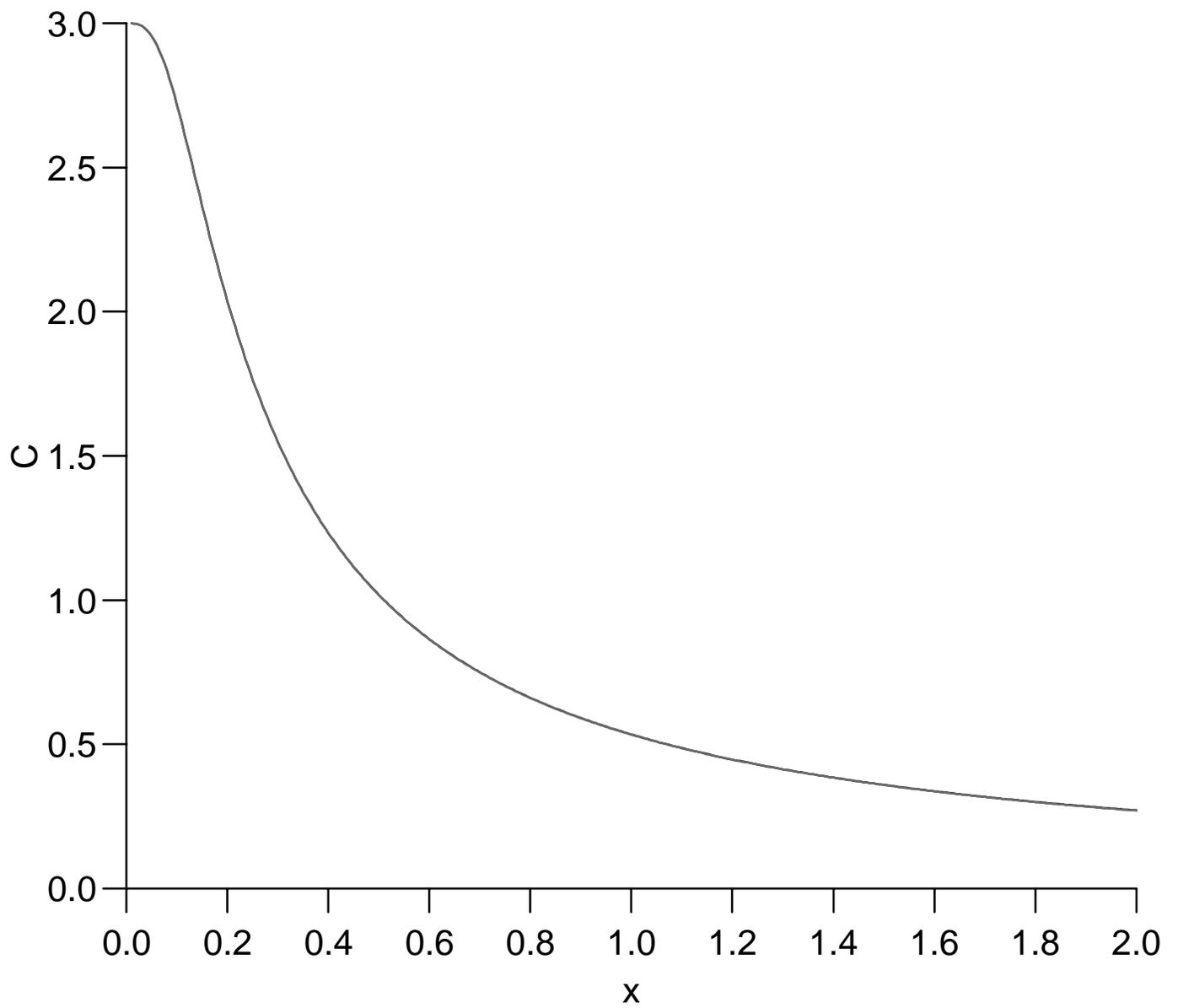
\begin{figure}
\caption{the familiar curve of the specific heat ploted against $x=\hbar
w/k_{B}T$ for $\lam =\lam '=1$ (isotropic harmonic oscilator) and
$N=1000$. Both the analytical and numerical results are shown, but they
are so similar that it looks like one curve only.}
\label{fig1}
\end{figure}

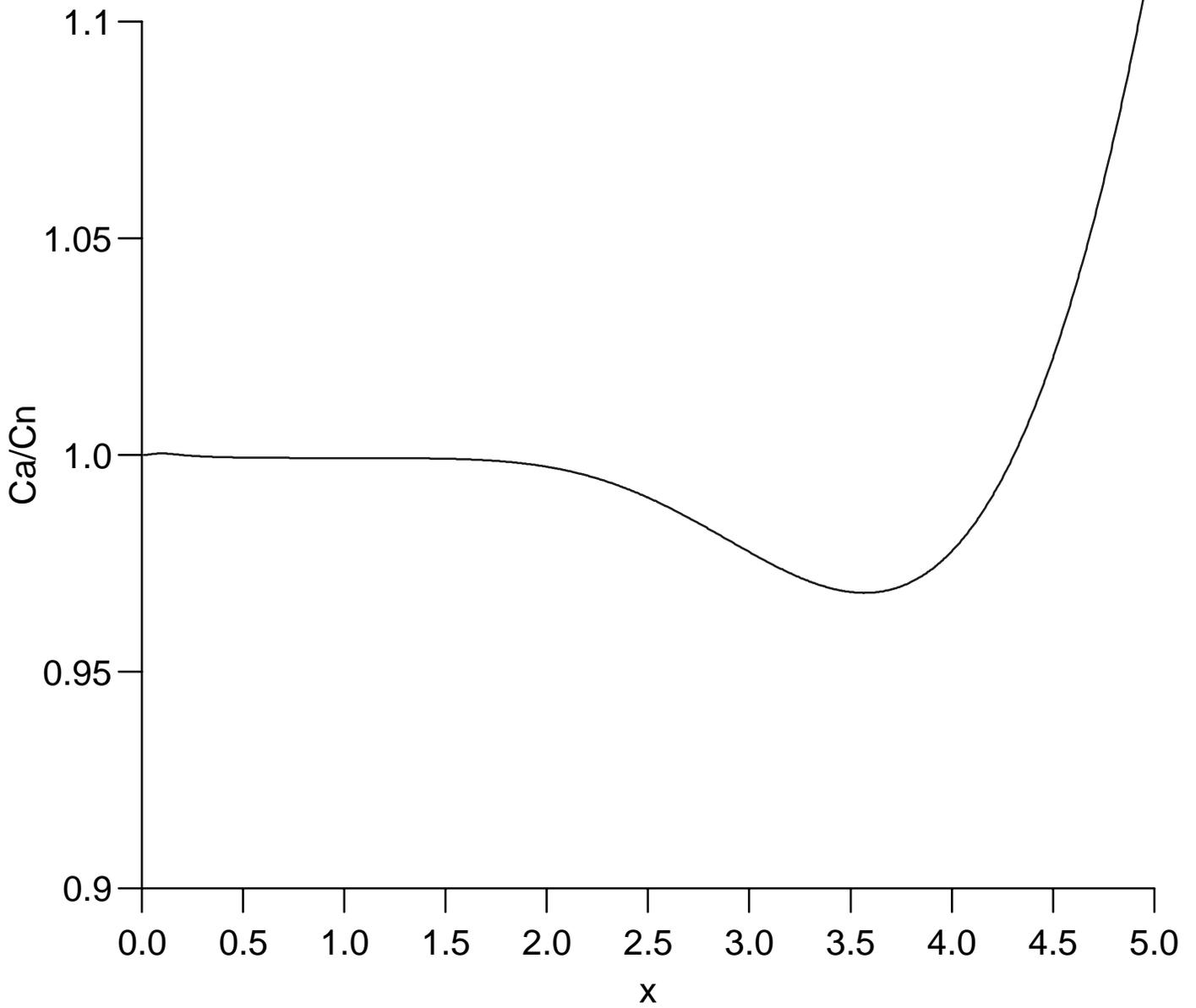
\begin{figure}
\caption{the ratio $C_{analytical}/C_{numerical}$ ploted against $x$ for
$\lam =\lam '=1$ and $N=1000$}
\label{fig2}
\end{figure}

\end{document}